\documentclass{iau} 
\usepackage{graphicx}
\usepackage{aas_macros}

\title[The Euclid Data Processing Challenges] 
{The Euclid Data Processing Challenges}

\author[Pierre Dubath et al.]   
{Pierre Dubath$^1$\footnote{On behalf of the Euclid collaboration},
Nikolaos Apostolakos$^1$,
Andrea Bonchi$^{2,3}$, 
Andrey Belikov$^{4}$, 
Massimo Brescia$^{5}$, 
Stefano Cavuoti$^{5,6}$, 
Peter Capak$^{7}$, 
Jean Coupon$^1$, 
Christophe Dabin$^{8}$, 
Hubert Degaudenzi$^1$, 
Shantanu Desai$^{9}$, 
Florian Dubath$^1$, 
Adriano Fontana$^{3}$, 
Sotiria Fotopoulou$^1$, 
Marco Frailis$^{10}$, 
Audrey Galametz$^{11}$, 
John Hoar$^{12}$, 
Mark Holliman$^{13}$, 
Ben Hoyle$^{11, 14}$, 
Patrick Hudelot$^{15}$, 
Olivier Ilbert$^{16}$, 
Martin Kuemmel$^{14}$, 
Martin Melchior$^{17}$, 
Yannick Mellier$^{15,18}$, 
Joe Mohr$^{14}$, 
Nicolas Morisset$^1$, 
St\'ephane Paltani$^1$, 
Roser Pello$^{19}$, 
Stefano Pilo$^{3}$, 
Gianluca Polenta$^{2,3}$, 
Maurice Poncet$^{8}$, 
Roberto Saglia$^{11}$, 
Mara Salvato$^{11}$, 
Marc Sauvage$^{20}$, 
Marc Schefer$^1$, 
Santiago Serrano$^{21}$, 
Marco Soldati$^{17}$, 
Andrea Tramacere$^1$, 
Rees Williams$^{22}$, 
Andrea Zacchei$^{10}$}

\affiliation{
$^1$Department of Astronomy, University of Geneva, Switzerland  
\\ email: {\tt Pierre.Dubath@unige.ch} 
\\[\affilskip]
$^{2}$ ASI Science Data Center, Roma, Italy 
\\[\affilskip]
$^{3}$ INAF, Osservatorio Astronomico di Roma, Monte Porzio Catone, Italy
\\[\affilskip]
$^{4}$ Kapteyn Astronomical Institute, University of Groningen, the Netherlands
\\[\affilskip]
$^{5}$ INAF, Astronomical Observatory of Capodimonte, Napoli, Italy
\\[\affilskip]
$^{6}$ Department of Physics, University Federico II, Napoli, Italy
\\[\affilskip]
$^{7}$ California Institute of Technology, Infrared Processing and Analysis Center, Pasadena, USA
 \\[\affilskip]
$^{8}$  Centre National d'\'Etudes Spatiales, Toulouse, France
 \\[\affilskip]
$^{9}$ Department of Physics, IIT Hyderabad, Kandi, India
\\[\affilskip]
$^{10}$ INAF - Osservatorio Astronomico di Trieste, Italy
\\[\affilskip]
$^{11}$ MPE, Max-Planck-Institut f\"ur Extraterrestrische Physik, Garching, Germany
\\[\affilskip]
$^{12}$ European Space Astronomy Centre, European Space Agency, Madrid, Spain
 \\[\affilskip]
$^{13}$ Institute for Astronomy, University of Edinburgh  
\\[\affilskip]
$^{14}$ Fakult\"at für Physik, Ludwig-Maximilians-Universit\"at, M\"unchen, Germany   
\\[\affilskip]
$^{15}$ CNRS-UPMC, Institut d'Astrophysique de Paris, France
\\[\affilskip]
$^{16}$ Aix Marseille Univ, CNRS, Laboratoire d'Astrophysique de Marseille (LAM), France 
\\[\affilskip]
$^{17}$ Fachhochschule Nordwestschweiz, Institute of 4D Technologies, Windisch, Switzerland  
\\[\affilskip]
$^{18}$ IRFU, Service d'Astrophysique, CEA Saclay, France
\\[\affilskip] 
$^{19}$ Universit\'e de Toulouse, UPS-OMP, CNRS, IRAP, France  
\\[\affilskip]
$^{20}$ Laboratoire AIM Paris-Saclay, CEA/IRFU/SAp, CNRS, Universite Paris Diderot, France
\\[\affilskip]
$^{21}$ Institute of Space Sciences (IEEC-CSIC), Campus UAB, Barcelona, Spain 
\\[\affilskip]
$^{22}$ Center of Information Technology, University of Groningen, the Netherlands 
}

\pubyear{2015}
\volume{325}  
\jname{Astroinformatics}
\editors{M. Brescia, et al. eds.}
\begin{document}

\maketitle

\begin{abstract}
Euclid is a Europe-led cosmology space mission dedicated to a visible and near infrared survey of the entire extra-galactic sky. Its purpose is to deepen our knowledge of the dark content of our Universe. After an overview of the Euclid mission and science, this contribution describes how the community is getting organized to face the data analysis challenges, both in software development and in operational data processing matters. It ends with a more specific account of some of the main contributions of the Swiss Science Data Center (SDC-CH). 
\keywords{dark matter, galaxies: distances and redshifts, galaxies: fundamental parameters, surveys, catalogs, methods: data analysis}
\end{abstract}

\firstsection 
              
\section{The Euclid Science and Mission}

The European Space Agency (ESA) announced the selection of the Euclid space mission in October 2011, at the same time as the Nobel Price in Physics was being awarded to Perlmutter, Schmidt, and Riess  ``for the discovery of the accelerating expansion of the Universe through observations of distant supernova''.  A connection between these two events can be outlined. The latter brings a wide recognition on the observational evidence for the existence of the so-called dark energy, while the former aims at deriving new constraints on the structure and the nature of the dark contents of our Universe. 

Despite huge progress accumulated in astronomy over the last decades, we still know close to nothing about the nature of the by far largest fraction of the matter - energy content of our Universe. The Euclid mission aims at improving our knowledge on the invisible stuff from detailed observations of the visible one, i.e., from shape and redshift determinations of a very large number of distant galaxies. Accurate analysis of the irregularities of these measurements can lead to stringent constraints on the properties of dark matter and dark energy.   Similar programs are already on-going with, e.g., the COSMOS survey (see e.g. 
\cite[Laigle et al.\,2016]{COSMOS2016}),
the Kilo Degree Survey (KiDS, see e.g. 
\cite[Hildebrandt et al.\,2016]{KIDS2016})
and the Dark Energy Survey (DES, see the 
\cite[DES Collaboration et al.\,2016]{DES2016Collaboration}).  
With a dedicated space telescope however,  the Euclid mission brings this type of survey to an unprecedented breadth, both in terms of accuracy and of sky coverage.

Euclid is a medium-class spacecraft mission of the Cosmic Vision program of ESA, planned for a 2020 launch on a Soyuz rocket from the European spaceport at Kourou (French Guiana) to the L2 Lagrangian point. The 1.2 meter main mirror telescope is equipped with an optical imager and a near-infrared instrument. The six-year nominal program includes both a 40 square degrees deep and a 15,000 square degrees wide surveys. The collected galaxy shape and redshift measurements can be analyzed  to characterize different cosmological effects (often called ``probes'') listed below.

\subsection{Weak Lensing}

\begin{figure}[h]
\begin{center}
 \includegraphics[width=3.5in]{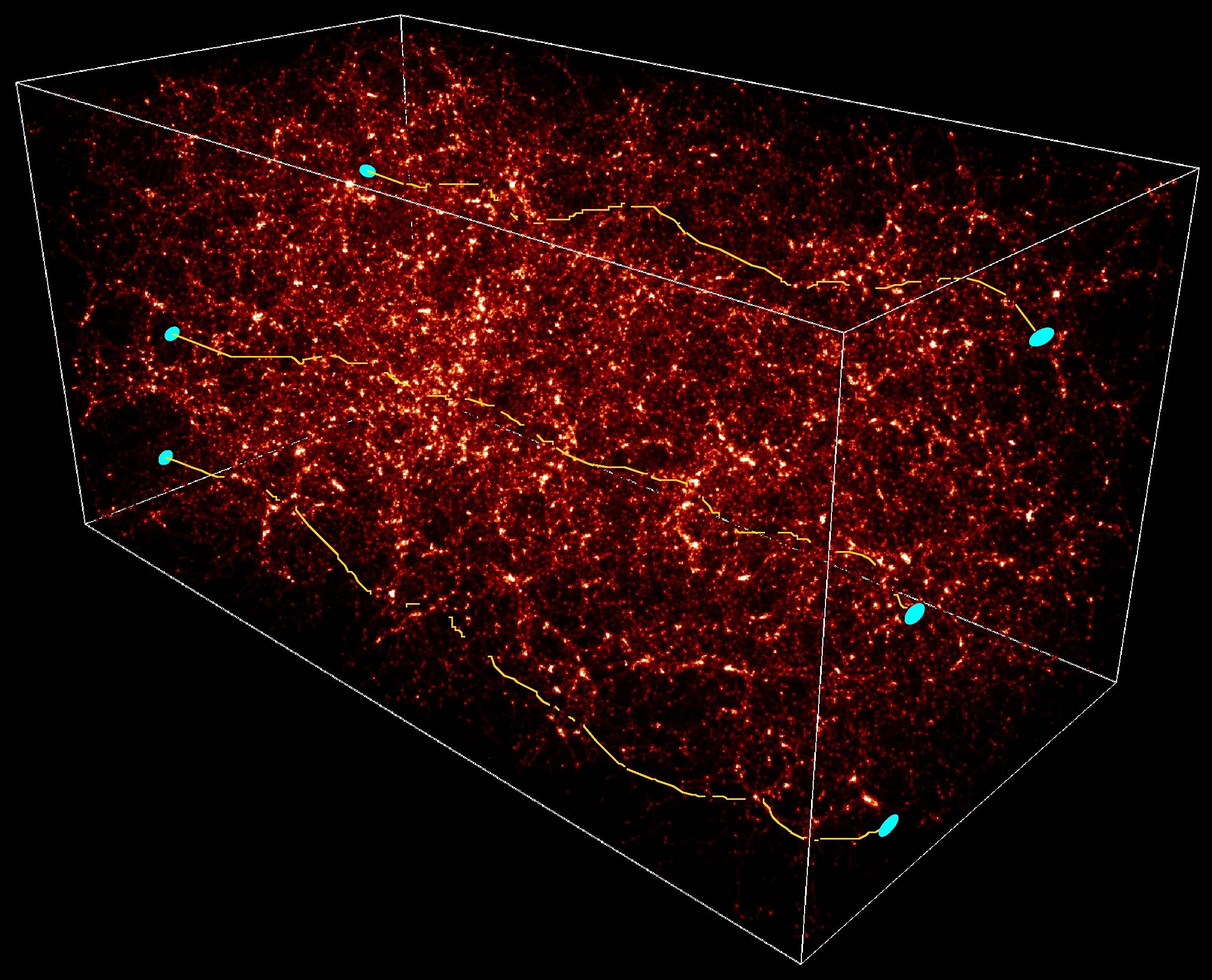} 
 \caption{Illustration of the deflection of three distant galaxy light rays crossing the Universe (Copyright CNRS/IAP/Colombi/Mellier) }
   \label{fig_weak_lensing}
\end{center}
\end{figure}

The  strong lens images attract much attention. Their large-scale arc-like structures result from lensing effects caused by the curvature of space around the huge central mass concentration. Weak lensing is an analogous, although less spectacular effect. As the light of the most distant galaxies travel through space to reach the Earth, it gets deflected by the mass concentrations crossed along the way. The alignment is most often imperfect, the mass concentration smaller and, as a consequence, the galaxy images are only slightly distorted.  Figure\,\ref{fig_weak_lensing} illustrates how the light of three distant galaxies gets deflected as it travels through a portion of Universe. The reddish dark matter structures shown in this figure result from cosmological simulations. 

Constraints on the total mass distribution can be derived from the knowledge of the shapes and the redshifts of all visible galaxies. The shape measurements are performed using sophisticated image analysis algorithms. They are however, affected by noise and also reflect the combination of intrinsic galaxy shapes and orientations. The distortions due to weak lensing alone can only be derived by averaging measurements from many different galaxies located in the same sky patch.  

\subsection{Baryonic Acoustic Oscillation (BAO)}

Oscillations analogous to sound waves were created by interactions between the gravity and pressure forces in the very hot and dense plasma of the early Universe.  The corresponding waves ``froze'' suddenly when protons and electrons combined to form the first atoms, freeing photons and canceling pressure in the process. Afterwards, under the influence of gravity, matter formed slowly increasingly denser regions, but the typical distance between over-densities  expressed in co-moving units have since not changed significantly. Measuring the typical separation between density peaks at different redshifts reveals how the Universe has expanded as a function of time. In this way, characterizing the BAO scale in the local Universe provides a direct measurement of the expansion acceleration completely independent from the distant supernovae observations. 

\subsection{Integrated Sachs-Wolfe (ISW) and redshift-space distortions (Kaiser effect) }

Other effects arise as perturbations of the galaxy apparent flow. One is the so-called Integrated Sachs-Wolfe (ISW) effect. When distant galaxy photons travel through space towards the observer, they may cross important mass concentrations formed by large galaxy clusters. They gain a little bit of energy when falling into the corresponding potential well and they would lose the same amount when going out on the other side. However, the Universe continues expanding while photons travel across the galaxy cluster, and escaping the well on the other side becomes a little bit easier. As a consequence, photons experience a net energy gain as they cross the cluster. Other redshift-space distortions are also caused by galaxy clusters. Inside the cluster, galaxies acquire large random velocities due to their mutual attraction. Outside, galaxies gain peculiar velocities as they tend to fall towards the cluster center. Cosmological information can be extracted from all these effects which will be better characterized through Euclid observations. 

\subsection{Legacy Science}

The Euclid mission data, with imaging of almost the full portion of sky outside the Galaxy plane at almost the Hubble telescope resolution and spectra for several tens of millions of galaxies will be an extremely valuable set for almost all types of extragalactic studies. 

\section{The Euclid instruments}

Along the optical path of the Euclid telescope, a dichroic plate splits the light collected by the 1.2 meter Korsch silicon carbide primary mirror into two channels leading to the VISible (VIS) and the Near Infrared Spectrometer and Photometer (NISP) instruments.  

The visible VIS instrument includes a mosaic of 36 e2v CCD detectors, each of them with 4096 by 4132 12 micron pixels. With a scale of 0.1 arcsec per pixel on the sky, the total field of view is 0.54 square degrees. The limiting magnitude through the unique photometric filter covering the 550 to 900 nm range is 24.5 in AB magnitude for a 10 sigma detection. VIS will provide high resolution visible images with a typical full-width at half-maximum of 0.23 arcsec. 

The  NISP near-infrared instrument comprises a mosaic of 16 Teledyne TIS H2RG detectors of 2040 by 2040 18 micron pixels. With a pixel size of 0.3 arcsec on the sky, the total field of view is 0.53 square degrees, i.e., almost identical to the VIS one, but with a much lower imaging resolution. The filter wheel assembly is equipped with three infrared filters in the Y, J and H bands as well as with four low resolution near infrared grisms (a blue one from 920 nm to 1250 nm, and three red ones from 1250 nm, to 1850 nm) for slit-less spectroscopy. The red grisms are identical, but they have different orientations for disentangling possible overlapping spectra of neighboring sources. The limiting magnitude through the infrared filters  is 24 in AB magnitude for a five sigma detection.

The volume of compressed detector data downloaded to the ground from the Euclid instruments  will be about 100 GB per day, i.e., a total of 200 TB after the six-year nominal mission duration. 

\section{Photometric redshift determination with ground photometry}

Good photometric measurements are expected to be collected for more than 1.5 billion of galaxies and quality spectra for up to 30 millions of them. Redshift measurements of the largest possible number of galaxies are required to reach the Euclid ambitious scientific goals. Accurate results will be derived from the infrared spectra, but for a vast majority of the galaxies, redshift measurements will be based on the photometry.  In few words, the more distant a galaxy, the redder it appears, as the photons get larger stretches resulting from the Universe expansion during their journey towards the observer. This basic effect is exploited to derive redshifts through the so-called photometric redshift algorithms, although the details are more complicated as galaxies have intrinsic colors and reddening can also be caused by dust absorption along the line of sight (see Sect.\,\ref{sect_phosphoros}).

A necessary condition to obtain good photometric redshifts is a sufficiently large number of filters, covering a wide wavelength range such as to provide stringent constraints on the underlying galaxy spectral energy distribution. The four Euclid bands are not enough. Imaging through additional filters from ground telescopes is absolutely necessary to the success of the Euclid mission. The ground photometry program is  acknowledged as an integral part of the mission. Agreements with the Dark Energy Survey (DES) and the Kilo-Degree Survey (KiDS) have been finalized, while negotiations with other ground surveys such as the Large Synoptic Survey Telescope (LSST) and with dedicated projects at the Javalahambre telescope in Spain and at the Subaru and the CFHT telescopes in Hawaii are at different levels of advancement.   

\section{The data analysis functional decomposition \label{sect_decomposition}}

\begin{figure}[h]
\begin{center}
 \includegraphics[width=5in]{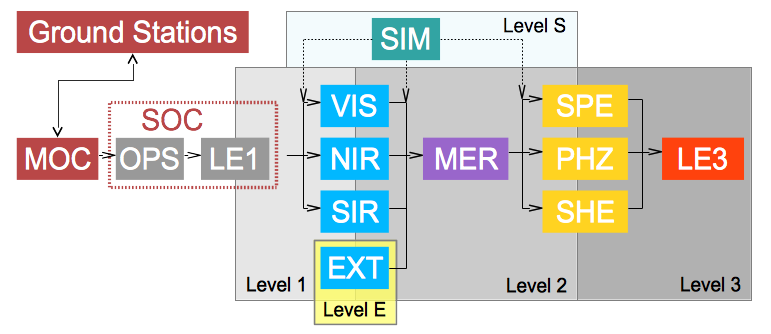} 
 \caption{Simplified picture of the Euclid data analysis functional decomposition}
   \label{fig_decomposition}
\end{center}
\end{figure}

A slightly simplified picture of the functional decomposition of the Euclid data analysis system is shown in Fig.\,\ref{fig_decomposition}. As often for space missions, ESA keeps the responsibility of the ground stations, of the Mission Operation Center (MOC) and of the Science Operation Center (SOC). These centers are in charge of the spacecraft operations, both from the up-link and the down-link.  Integrating help provided by the Euclid consortium, one of the SOC task is to decode the telemetry packets and store the raw data in the form of computer readable numbers. These ``level 1" raw data are the starting point of the functional decomposition.

The low-level cleaning and corrections, such as bias subtraction, flat fielding, cosmic ray removal and charge transfer inefficiency corrections of the CCD frames as well as their astrometric and photometric calibration are divided into different tasks according to the data type. They are grouped into the VIS, NIR, SIR and EXT processing functions for the visible image, the near infrared image, the near infrared spectra and the ground CCD frames, respectively.

The task of the MER processing function is to merge the outputs of VIS/NIR/SIR processing together with EXT corresponding results. The main goal is to produce the basis of the Euclid catalog with source identifications, calibrated flux measurements and spectra. As some of the sources may exhibit a significant signal only in some of the photometric bands, the final source detection can only happen at the MER level, when all CCD frames can be analyzed together. But all data have to be put on a common ground before, in particular from the astrometry point of view. 

The SPE and PHZ processing function tasks consist of deriving the spectroscopic and the photometric redshift measurements, while SHE is concerned with the galaxy shape determinations. The strong lenses detection is also part of SHE, but it has a lower priority and visibility. It is considered a secondary science topic. Finally, the high-level LE3 processing produces the cosmological parameter measurements which will be part of the final Euclid catalog. The SIM processing function produces different levels of simulated data, as required to support the integration tests performed before the launch.

\section{Processing budget estimations}

The total amount of data coming out of the Euclid spacecraft is relatively modest by today's big-data standards. The 200 TB of compressed telemetry data will, however, be expanded by a large factor (several tens) in the course of the data analysis. The volume of the ground photometric data is also expected to be large, current estimations amount to about 10 PB. There are also large numbers expected from LE3 simulations related to forwards modeling of the data. All in all the total amount of data is anticipated to raise to some tens of PB with 100 PB as a plausible upper limit. Estimation of the amount of CPU time required to achieve the Euclid data processing tasks is a very daunting task as  in many cases, the final details of the algorithms are not yet known. Numbers presented at a recent ESA review suggest that a computer cluster equipped with 20,000 processing cores of typical 2016 performance dedicated full-time to the task would be able to achieve all required processing.

\section{Science ground segment organization \label{sect_sgs}}

\begin{figure}[h]
\begin{center}
 \includegraphics[width=5.2in]{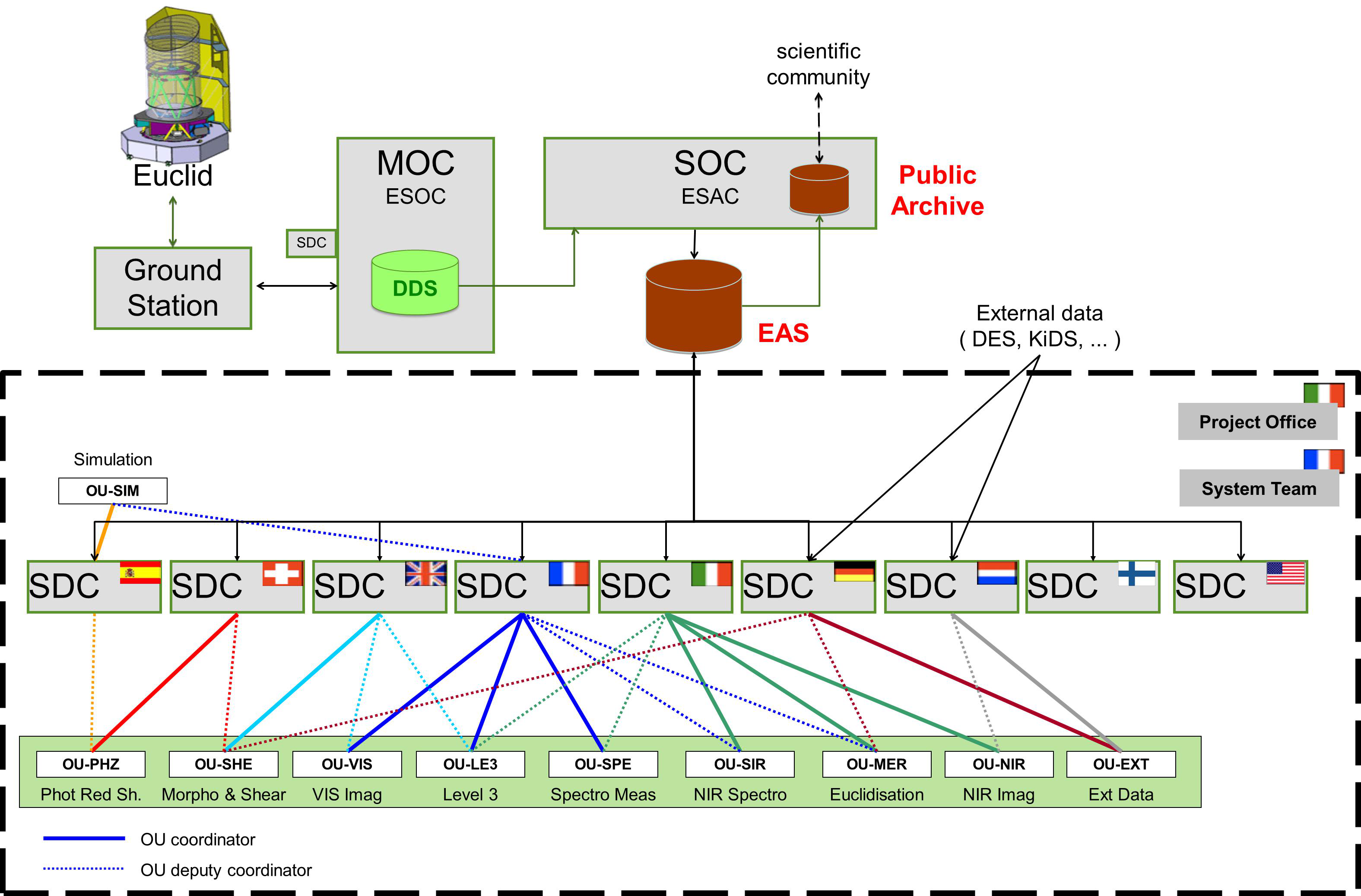} 
 \caption{Sketch of the 2016 Euclid Science Ground Segment (SGS) organization}
   \label{fig_sgs}
\end{center}
\end{figure}

In order to distribute the workload related to both software development and data processing after the launch, the Euclid community organized itself into a number of groups. One Science Data Center (SDC) team was created in each of the participating countries, while an Organizational Unit (OU) consortium was formed around each of the processing functions described in Sect.\,\ref{sect_decomposition}. Figure\,\ref{fig_sgs} displays a sketch of the Science Ground Segment (SGS) organization (modeled on that of the Gaia mission established a few years ago).

The task distribution among the different types of groups is rather straightforward. Scientists organized in Science Working Groups (SWGs) are in charged of setting up the scientific requirements.  The OUs are responsible for studying  possible solutions, i.e., finding algorithms meeting the requirements given the properties of the Euclid data. And the SDCs are developing and integrating the corresponding software, following accepted common standards. The SDCs are also in charge of the data processing after the launch, making sure the required computer infrastructure will be available. The so-called System Team (ST), also shown in Fig.\,\ref{fig_sgs}, includes representatives of all SDCs. It is in charge of coming up with solutions for all commonalities across SDCs. It is also responsible for establishing the coding standards and for the development of common tools, such as libraries and the components of the distributed processing system (see below Sect.\,\ref{sect_dist_proc}).



\section{Software development}

 Minimizing the number of solutions introduced into a data processing system is essential to limit its complexity. The most important constraints in the Euclid case are the usage of Python and of the C++ languages on a reference platform of the Red Hat Linux family. The corresponding standard for software packaging and installation is the RPM package manager. Different virtualization solutions are being tested, but a final choice has not yet been made.

The list of accepted external libraries is controlled through a formal procedure called EDEN. The full software development environment is made available in a virtual machine (LODEEN). All source code is committed on a GitLab common repository and continuous integration is performed on a centralized Jenkins-based  system (CODEEN). 

An XML-based common data model is in development. It is probably the most important element of the collaboration as it represents the concrete interfaces between all software components. Finally, a common building and packaging framework named Elements is proposed to the entire community.

\subsection{The Elements Framework \label{sect_elements}}

Elements is a C++ and Python CMake-based building and packaging framework, originally derived from the CERN Gaudi project (see http://proj-gaudi.web.cern.ch/proj-gaudi/). Software projects are organized into a number of independent modules conforming to a standard repository structure. The building instructions are provided through CMakeList.txt files and underlying CMake processes automatically generate usual Makefiles. The Elements framework also provides a number of basic services, such as standard solutions for logging and program options handling. One of the main strengths of Elements is the ability to manage different software versions, for programs and their full hierarchy of dependencies, on a  single system. Multiple versions are installed on custom locations, e.g., on ``/opt/.../{\it project\_name}/{\it project\_version}/.../bin/{\it an\_executable}'' rather than the Linux standard /usr/bin/{\it an\_executable}. With the project version number being part of the path, different versions can be installed in parallel and Elements ensures that both build and run-time environments are correctly setup (through an automatic setting of environment variables such as PATH, LD\_LIBRARY\_PATH and PYTHONPATH). 

\section{Distributed Data Processing \label{sect_dist_proc}}

The Euclid data processing load of the operational phase is to be divided into all SDCs. Different models can be envisaged. SDCs could process the data in turn through different analysis levels, forming a processing chain where each SDC takes inputs from the previous step and feeds outputs to the next one. At the other extreme, each SDC could run the entire processing pipeline from the raw data to the final results. The model currently tested for Euclid is close to the second one. Most of the analysis tasks are expected to run in parallel in the different SDCs. Exceptions may occur, however, with some specific tasks being executed on dedicated SDCs (e.g., for producing calibration results needed as inputs into the main pipeline).

\begin{figure}[h]
\begin{center}
 \includegraphics[width=5in]{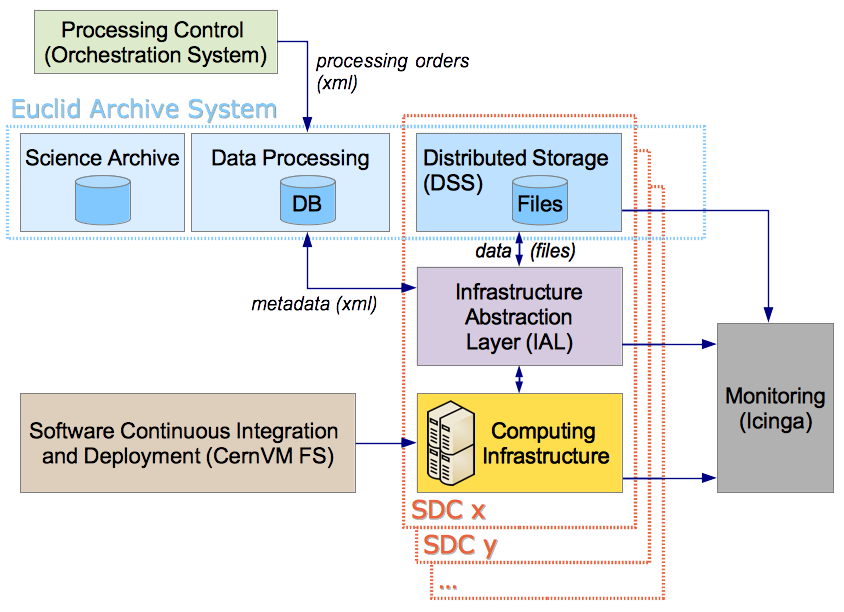} 
 \caption{Schematic view of the Euclid distributed processing system}
   \label{fig_dist_proc}
\end{center}
\end{figure}

Figure\,\ref{fig_dist_proc} shows a sketch of the proposed distributed processing system. A meta-data database (with a remote duplicate for security reasons) lies at the heart of the system. It contains information about the data files and the status of the processing. A centralized control tool is used to create {\it processing orders} and to ingest them in the database. Infrastructure Abstraction Layer (IAL) processes running in each SDC query the database for processing orders. When the IAL running on a given SDC gets a processing order, it first checks if all required input data are available locally. It interacts with the distributed Storage System (DSS) to copy data files from other SDCs if necessary, before submitting corresponding jobs to the local computing cluster infrastructure. 

The software deployment is ensured by the CernVM file system, which mounts on each cluster node a remote directory containing an installation of the required software. Another tool is being setup to monitor the overall status of the processing across all SDCs. The full Euclid archive system, including data from any processing level, is distributed over all SDCs and the DSS system makes sure that copies of any data set are at least available from two different SDCs (for security and data transfer efficiency reasons). 

\section{Challenge driven development}

The development schedule of the Euclid data processing system is organized around large-scale integration test challenges, both for the scientific data analysis pipelines and for the more technical elements of the distributed processing system. This challenge-driven approach has proven to be an efficient way of progressing, despite the complex SGS organization. It follows an incremental and iterative test-driven path and it consolidates the interfaces, through concrete utilization of the common data model. The challenges conducted in 2016 allowed to test preliminary versions of all different components of the distributed processing system (see Sect.\,\ref{sect_dist_proc}). The resulting infrastructure was also used to run pipelines containing the VIS, NIR and SIR related data analysis software\footnote{Simulated data produced by the SIM team have been used as input data since the very first challenges achieved in 2015.}. The coming 2017 challenge plans to add the SIM,  EXT and MER processing functions to the picture, running on improved versions of the underlying infrastructure distributed-processing components. The current Euclid schedule includes a series of ever more ambitious challenges that will gradually build up the entire Euclid data processing system.

\section{Some Swiss SDC contributions \label{sect_sdcch}}

The SDC-CH is developing the Elements building and packaging framework described in Sect.\,\ref{sect_elements}. The photometric-redshift determination pipeline is another task of the SDC-CH. Even if different options continue to be tested, an overall scheme is emerging. It starts with a classification aiming at separating stars, galaxies and AGN. The next step involves different photometric redshift determination methods (from template fitting and machine learning algorithms) applied in parallel and optimized for different regions of the input parameter space. The optimum combination of the results is then achieved with another machine learning classifier, trained with a spectroscopic redshift sample. For each cell of the input color space, a further bias correction step is applied by shifting the photometric redshift distribution mean value to that of the spectroscopic redshifts falling on the same color cell. 

\subsection{Phosphoros \label{sect_phosphoros}}

Phosphoros is a new C++ implementation of a photometric-redshift template fitting algorithm. Galaxy spectra are modeled by red-shifting and reddening template Spectral Energy Distributions (SED). The sets of SEDs, of redshift and reddening steps considered form a three-dimensional parameter space (extra dimensions can be added in advanced analyses). Modeled photometric values can be computed for each cell of this parameter space by integrating the spectra through the filters and the values can be compared to the observed ones. A model-to-observation distance can be measured (through a $\chi^2$ calculation for example) and a likelihood derived, again for each cell. The highest peak in the  multi-dimensional parameter space indicates the most likely model. To turn this scheme into a full Bayesian one, the likelihoods are multiplied by priors and values along unwanted axes are marginalized to produce one-dimensional Probability Density Functions (PDF). The main PDF gives probabilities as a function of redshift values, but PDFs can also be obtained along other axis, for determinations of physical parameters, such as galaxy mass or star formation history. Phosphoros can take as input priors along any axis, including luminosity (luminosity priors can be specified as luminosity functions). It currently also includes options to add to the relevant template SEDs emission lines  and it  implements a new scheme for handling galactic extinction (\cite[Galametz et al.\,2016]{GalacticExtinction2016}). Phosphoros is expected to be further developed and maintained throughout the full Euclid mission lifetime.

\subsection{SExtractor++ development \label{sect_sex}}

SExtractor (
\cite[Bertin \& Arnouts 1996]{Sextractor})
is one of the most successful astronomical software tools. A couple of years ago, an assessment of the long-term update and maintenance plans, made clear that time was up to start the development of a new SExtractor implementation. A C++ project was initiated to provide a useful tool to both the Euclid collaboration and the astronomical community.  SDC-CH members bring modern expertise in modular object-oriented architectural design. A single responsibility is attributed to each code element, software interfaces are used in a systematic manner and many design patterns are at the heart of the new project. A modular framework (together with aperture photometry functionalities) is now in place, a preliminary proof-of-concept version of multi-frame model fitting has been tested while the current focus is on the development of a convenient configuration system. A first release of the new SExtractor project is planned for 2017.

\subsection{Strong Lens Detection}

A collaboration with scientists from colleagues from the SHE processing function (see Fig.\,\ref{fig_sgs})  is ongoing to tests different algorithms of source (including strong lens) detection and deblending (see 
 \cite[Paraficz et al.\,2016]{Danka2016} and
\cite [Tramacere et al.\,2016]{Asterism2016}).

\section{Conclusion}

To conclude, let us come back to the word ``challenge'' used in the title of this contribution. There are challenges everywhere in the Euclid mission. Very ambitious scientific goals push some of the hardware requirements to their limits and trigger quests for perfect analysis algorithms and related software implementations. But human factors should not be overlooked. The international community is organizing itself in the absence of a contract-based hierarchy both for software development and for operation preparation and this is the real ``challenge''. Fortunately, the Euclid challenge-driven approach has proven to be an amazingly efficient method for steering large collaborations towards common goals. Capitalizing on the recent successes, the challenge-based current planning provides a promising path to complete the full analysis system on time, before the Euclid launch.

%
%


\end{document}